\def\arcmin{\hbox{$^\prime$}}
\def\farcs{\hbox{$.\!\!^{\prime\prime}$}}
\def\arcsec{\hbox{$^{\prime\prime}$}}

\documentclass{PoS}

\title{High resolution studies of Broad Absorption Line radio-loud quasars}

\ShortTitle{BAL quasars}

\author{\speaker{Marcin P. Gawro\'nski}
        \\
        Toru\'n Centre for Astronomy, N. Copernicus University, Gagarina 11, 87-100 Toru\'n, Poland\\
        E-mail: \email{motylek@astro.uni.torun.pl}}

\author{Magdalena Kunert-Bajraszewska\\
        Toru\'n Centre for Astronomy, N. Copernicus University, Gagarina 11, 87-100 Toru\'n, Poland\\
        E-mail: \email{magda@astro.uni.torun.pl}}

\abstract{Broad absorption lines (BALs), seen in a small fraction of both the
radio-quiet and radio-loud quasar populations, are probably caused
by the outflow of gas with high velocities and are part of the accretion
process. The presence of BALs is the geometrical effect and/or it is
connected with the quasar evolution.
Using the final release of FIRST survey combined with a A Catalog of BAL
QSOs (SDSS/DR3), we have constructed a new sample of compact radio-loud BAL
QSOs, which makes the majority of radio-loud BAL QSOs.
The main goal of this project is to study the origin of BALs
by analysis the BAL QSOs radio morphology, their orientation
and jets evolution, using EVN at 1.6\,GHz and VLBA at 5 and 8.4\,GHz.
We will discuss also the first multi-frequency radio observations of
very compact BAL quasar 1045+352 made
using the VLBA, EVN, MERLIN and CHANDRA. We speculate that the most probable
interpretation of the observed radio structure of 1045+352 is the ongoing
process of the jet precession due to internal instabilities within the flow.}

\FullConference{10th European VLBI Network Symposium and EVN Users Meeting: VLBI and the new 
generation of radio arrays\\
		September 20-24, 2010\\
		Manchester Uk}

\begin{document}

\section{Introduction}
Broad absorption lines (BALs) - high ionization resonant lines (C\,IV 1549$\AA$) and low ionization lines (Mg\,II
2800$\AA$), are seen in a small fraction of both the radio-quiet and radio-loud quasar populations (10--30\%, 
depends on selection criteria, e.g \cite{jia07}) . It is suggested that the presence of BALs is a pure
geometrical effect \cite{elv00} or/and is connected with the quasar evolution scheme \cite{bec00}.

BALQSOs are identified by means of their Balnicity Index $BI>0$ \cite{wey91}
but very recently a less restrictive criterion, Absorption Index $AI>0$
\cite{tru06}, has been defined and used. $BI$ is computed
considering any absorption trough spanning $\geq 2000~{\rm km~s^{-1}}$ in
width, absorbing at least 10\% of the local continuum, and blueshifted by
$\geq 3000~{\rm km~s^{-1}}$ with respect to the corresponding emission
lines. The Absorption Index is computed in the same way as $BI$, but
relaxing the ${\rm 3000~km~s^{-1}}$ blueshift criterion and considering all
absorption troughs with a blueshift $>0~{\rm km~s^{-1}}$ with respect to the
corresponding emision lines, and with a width of at least 1000 ${\rm
km~s^{-1}}$.
The validity of the latter definition has been recently questioned (e.g. \cite{gan07}). The BALQSOs 
are classified into three subclasses: High-ionization BALs (HiBALs) contain strong, broad absorption high-ionization 
lines and are typicaly identified through the presence of CIV absorption; Low-ionization BALs (LoBALs) contain 
HIBAL features but also have absorption from low-ionization lines (MgII) present in optical spectra; FeLoBALs are 
LoBALs with exited-state FeII and FeIII absorption. 

Theoretical models (\cite{elv00},\cite{mur95}) suggest that BALs are seen at high inclination angles, which means that 
the outflows from accretion disks are present near an equatorial plane. However, some recent numerical work indicates
that it is also plausible to launch bipolar outflows from the inner regions of a thin disk (e.g. \cite{pro04}).
There is a growing observational evidence indicating the existence of polar BAL outflows (e.g. \cite{zho06}).
This means that there is no one simple orientation model which can explain all the features observed in BAL quasars.

The radio morphologies of radio-loud BAL quasars provide important additional information about their orientation 
and the direction of the outflow. Currently, there are only 17 compact radio-loud BAL quasars observed with VLBI in 
the literature (e.g. \cite{kun10}, \cite{liu08}). About half of them have still unresolved radio structures even in 
the high resolution observations, the other have core-jet structures indicating some re-orientation or very complex 
morphology, suggesting a strong interaction with the surrounding medium
\cite{kun10}. Recent observations support the 
evolutionary scheme \cite{mon09}. 

\section{Observations of BAL quasars}
\subsection{BAL quasar 1045+352 - summary}

1045+352 is a compact radio-loud CSS source.  The spectral
observations \cite{wil02} have shown 1045+352 to be a quasar with
a redshift of $z=1.604$. It has also been classified as a HiBALQSO   
based upon the observed very broad C\,IV absorption, and it is a very
luminous submillimetre object with detections at both 850\,$\mu$m and
450\,$\mu$m \cite{wil02}.  The radio luminosity of 1045+352 at
1.4\,GHz is high (log$L_{1.4\mathrm{GHz}}$=28.25~W~${\rm Hz^{-1}}$), making
this source one of the most radio-luminous BAL quasars.  
The Chandra X-ray observation of 1045+352
\cite{kun09} show that the X-ray emission of 1045+352 is very weak
in comparison to the other radio-loud BAL quasars which, together with
the high value of the optical$-$X-ray index $\alpha_{ox}=1.88$,
suggests a presence of an X-ray absorber close/in the BLR region.  The
result of the SED modeling of 1045+352 indicates the X-ray emission we
observe from 1045+352 can be mostly due to X-ray emission from the  
relativistic jet, while the X-ray emission from the corona is absorbed
in a large part. 

We have performed sensitive high-resolution radio observations of
the BAL quasar 1045+352 using the EVN+MERLIN at 5 GHz \cite{kun10}. The
radio morphology of 1045+352 is dominated by the
strong radio jet resolved into many sub-components and changing the
orientation during propagation in the central regions of the host
galaxy. The presence of the dense environment and material outflows can have
a significant impact on the radio jet morphology of 1045+352. We concluded that the most  
probable interpretation of the observed radio structure of 1045+352 is
the ongoing process of the jet precession due to internal
instabilities within the flow and/or jet interaction with a dense,
inhomogeneous interstellar medium.

\begin{table}[t]
\centering
\begin{tabular}{c|c|c|c|c|c|c }
\hline
  &  &  & Absorption & $F_{1.4\,GHz}$ &  & Radio \\
 Source & z & Type &  index /AI/ & /mJy/ & $\alpha^{1.4}_{4.8}$ & morphology \\
 (1) & (2) & (3) & (4) & (5) & (6) & (7) \\
\hline
0214-011  & 2.462 & HiBAL  & 791 & 218.0 & 0.60 & core-jet \\
0753+373$^{*}$   & 2.514 & HiBAL  & 841 & 247.3 & 0.04 & core-jet \\
0756+406$^{*}$  & 2.021 & HiBAL  & 1114& 199.6 & 0.87 & core-jet \\
0812+332$^{*}$  & 2.426 & nHiBAL & 747 & 342.5 & 0.94 & core-jet/double \\
0925+450A$^{*}$ & 1.904 & nHiBAL  & 293 & 162.2 & -0.35  & core-jet? \\
1002+483$^{*}$  & 2.372 & HiBAL  & 783 & 209.3 & 0.89 & core-jet \\
1010+495$^{*}$  & 2.201 & nHiBAL  & 361 & 269.4 & 0.39 & core-jet/double \\
1015+057  & 1.938 & HiBAL  & 441 & 296.6 & -0.09 & core-jet? \\
1040+080  & 2.665 & HiBAL  & 1011 & 381.6 & 0.71 & double\\
1054+036  & 2.832 & HiBAL  & 440 & 157.4 & 0.66 & core-jet \\
J1103+0232& 2.514 & HiBAL  & 460 & 166.0 & 0.34 & core-jet? \\
1157+014  & 2.000 & HiBAL  & 2887 & 268.5 & 0.64 & unresolved \\
1221+509$^{*}$  & 3.488 & HiBAL  & 413 & 228.8 & 0.59 & core-jet \\
1403+411$^{*}$  & 1.993 & HiBAL  & 780 & 214.0 & -0.21 & core-jet \\
1430+412$^{*}$  & 1.970 & nHiBAL  & 343 & 261.7 & 0.76 & complex\\
1526+533$^{*}$  & 2.822 & HiBAL  & 1701 & 182.6 & 0.71 & core-jet/double\\
\hline 
\end{tabular}
\caption{(2) based on SDSS; (3)-(4) taken from \cite{tru06}; (5) taken from FIRST; (6) $S\propto \nu^{-\alpha}$, 
based on FIRST and GB6; (7) radio morphology based on our observations and
VIPS images \cite{hel07}; ($^{*}$) also with EVN observations at 1.6\,GHz.}
\end{table}

\begin{figure}[t]
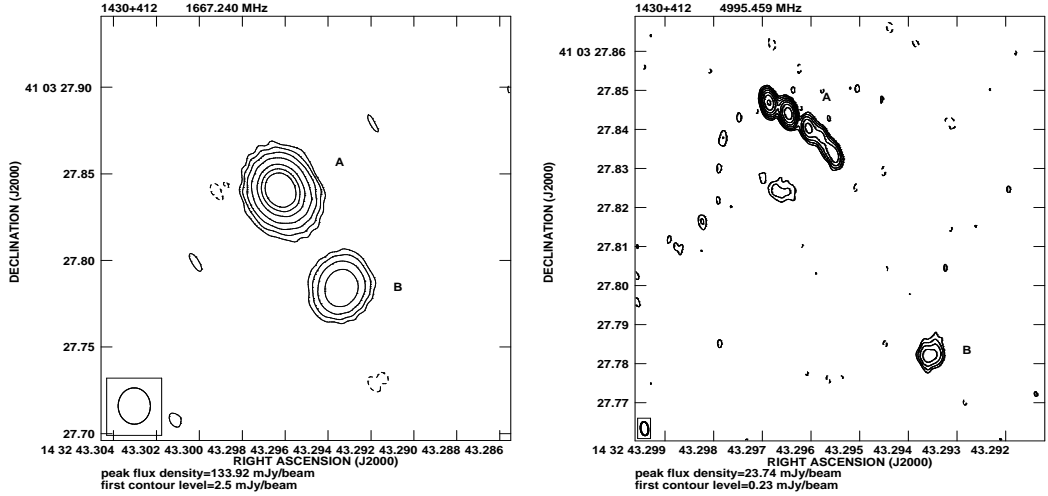

\hskip .0cm
\includegraphics[width=7cm, height=7cm]{1430+412.18CM.PS}
\includegraphics[width=7cm, height=7cm]{1430+412.6CM.PS}
\caption{
The EVN L-band (left) and VLBA C-band (right) maps of the BAL quasar
1430+412.}
\label{complex_maps}
\end{figure}

\subsection{Radio observations of a new sample of BAL quasars - preliminary
results}

We have prepared a project of multi-frequency radio observations of a new
sample of compact radio-loud BAL quasars.The main 
goal of this project was to study the origin of BALs, jets evolution and
probable jet-cloud interactions,  
by analysis of the BAL QSOs radio morphology. The high resolution radio
observations allow to resolve their
compact structures and to estimate the orientation of the jet axis what may
be a critical parameter
in the case of the BAL origin scenarios.

Using the final release of FIRST survey \cite{whi97} we looked for unresolved, isolated sources i.e. more 
compact than FIRST beam (5\farcs4) and surrounded by an empty field (we adopted 1\arcmin ~as a radius of 
that field). We set 150\,mJy as the lower flux density limit for the objects at 1.4\,GHz. We did not put any 
limits in the case of spectral index, because both GPS and CSS sources can be compact BAL quasars. 
The next step in our project was a cross-identification of the selected
FIRST objects with the optically selected
BAL quasars from {\it A Catalog of 
Broad Absorption Line Quasars from the Sloan Digital Sky Survey Third Data
Release} \cite{tru06}.
An automated algorithm was used to match the FIRST coordinates with the SDSS
position in a radius of 10\arcsec.
Finally, our sample consists of 16 sources (Table 1). 

We have observed the sample with VLBA at 5 and 8.4\,GHz with polarization detection on 27 \& 30 Aug 2008, 
6 \& 7 Sep 2008. The next part of survey was undertaken using EVN at 1.6\,GHz and the observations
were performed for 10 out of 16 target sources on 27 Oct 2008. 
However, we have noticed that there is an overlap between our sample of  
compact BAL quasars and recently published VIPS (VLBA Imaging and
Polarimetry Survey) objects \cite{hel07}. The authors obtained high dynamic range radio maps and 
polarization information for 5 of our sources at 5\,GHz. We have continued the observations of
our sample in a similar observational strategy to get comparable quality of the radio images for 11 
target sources at 5\,GHz and 16 objects at 8.4\,GHz. The target source scan was interleaved with 
a scan on a phase reference source
and the total cycle time (target and phase-reference) was
8 minutes including telescope drive times, with 6\,minutes actually on the target source per cycle.
The whole data reduction process was carried out using standard AIPS
procedures. For the target source, the corresponding
phase-reference source was mapped and the phase errors so determined were
applied to the target source, which were then mapped using a few cycles of
phase self-calibration and imaging. IMAGR was used to produce the final
radio images.

All target sources were detected, but they were finally resolved during  
observations at higher frequencies: 5 and 8.4\,GHz which means they are
generally very compact sources. Some of them show very complex morphology (Fig. 2 as an example) with probably fading
components suggesting they can be manifestation of the previous activity. This can be explained as caused
by the jet-ISM interactions or precession of the jet axis (see also work on the CSS BAL quasar 1045+352 by \cite{kun10}). 
Most of the observed sources revealed core-jet structures and show fainter emission from the jet compared to the
core. This morphology suggest intermediate orientation between polar and equatorial geometries    
and support the orientation scenario. For some sources the observed flux densities of compact structures account
only up to 25\% of total flux density at 5\,GHz (on the contrary to $\sim\,80\%$ in the case of sources from \cite{liu08}). 
This may suggest that there are low brightness extended structures in our new selected radio-load BALQSOs 
and sources may be older and bigger than GPS/CSS objects.

It should be noted here, that there are two important observational biases
in our sample. First, the selection criteria caused that all sources from
the new sample are HiBALs with absorption index AI\,$>$0 and balnicity index BI\,=\,0. Second, because of the SDSS
properties the automated algorithms used by \cite{tru06} could identify HiBALs via CIV from redshift range 
$1.7\leq z\leq 4.38$, therefore the  
selected sources are probably the most luminous radio-loud BALQSOs. So it is probable that we are probing 
a special group of radio-loud BAL quasars. 

A detailed discussion about the radio properties of the new sample and analysis of their spectral features will be presented 
in a forthcoming paper.

\bigskip
\noindent
{\footnotesize
{\bf Acknowledgement}\\
\noindent
MERLIN is a UK National Facility operated by the University of Manchester on behalf of STFC. The European VLBI Network is a joint
facility of European, Chinese, South African, and other radio astronomy institutes funded by their national research councils.
This work was supported by the Polish Ministry of Science and Higher Education under grant N N203 303635.

\end{document}